\documentstyle[preprint,aps,epsf]{revtex}

\draft

\begin{document}

\title{Hyperon-hyperon interactions and properties of 
       neutron star matter}

\author{I.\ Vida\~na, A.\ Polls and A.\ Ramos}

\address{Departament d'Estructura i Constituents de la Mat\`eria,
         Universitat de Barcelona, E-08028 Barcelona, Spain}

\author{L.\ Engvik and M.\  Hjorth-Jensen}

\address{Department of Physics, University of Oslo, N-0316 Oslo, Norway}

\maketitle

\begin{abstract}

We present results from Brueckner--Hartree--Fock calculations
for $\beta$-stable neutron star
matter with nucleonic and
hyperonic degrees of freedom, employing the most recent parametrizations
of the baryon-baryon interaction of the Nijmegen group.
It is found that the only strange baryons emerging in $\beta$-stable matter
up to total baryonic densities of 1.2 fm$^{-3}$ are $\Sigma^-$ and
$\Lambda$.
The corresponding equations of state are then used to compute properties 
of neutron stars such as masses and radii.

\end{abstract}



\section{Introduction}

The physics of compact objects like neutron stars offers
an intriguing interplay between nuclear processes  and
astrophysical observables.
Neutron stars exhibit conditions far from those
encountered on earth; typically, expected densities $\rho$
of a neutron star interior are of the
order of $10^3$ or more times the 'neutron drip' density 
$\approx 4\cdot 10^{11}$ g/cm$^{3}$,
where  nuclei begin to
dissolve and merge together.
Thus, the determination of an equation of state (EoS)
for dense matter is essential to calculations of neutron
star properties. The EoS determines properties  such as
the mass range, the mass-radius relationship, the crust
thickness and the cooling rate.
The same EoS is also crucial
in calculating the energy released in a supernova explosion.

At densities near to the saturation density of nuclear
matter, ($\rho_0=0.16$ fm$^{-3}$), 
we expect the matter to be composed of mainly neutrons, protons 
and electrons in $\beta$-equilibrium, since neutrinos have on average a
mean free path larger than the radius of the neutron star. The
equilibrium conditions can then be summarized as
\begin{equation}
    \mu_n=\mu_p+\mu_e,  \hspace{1cm} \rho_p = \rho_e,
     \label{eq:npebetaequilibrium}
\end{equation}
where $\mu_i$ and $\rho_i$ refer to the chemical potential and density
of particle species $i$, respectively.
At the saturation density of nuclear matter, $\rho_0$,
the electron chemical potential is
of the order $\sim 100$ MeV.
Once the rest mass of the muon is exceeded, it becomes
energetically favorable for an electron at the top
of the $e^-$ Fermi surface to decay into a
$\mu^-$. A Fermi sea of degenerate negative muons starts then to develop
and, consequently, the charge balance needs to be changed
according to $\rho_p = \rho_e+\rho_{\mu}$
as well as requiring that $\mu_e = \mu_{\mu}$.

As the density increases, new hadronic degrees of freedom may 
appear in addition
to neutrons and protons.
One such degree of freedom is hyperons, baryons with a
strangeness content.
Contrary to terrestrial conditions where hyperons are unstable and decay
into nucleons through the weak interaction, the equilibrium conditions
in neutron stars can make the inverse process happen, so that the
formation of hyperons becomes energetically favorable.     
As soon as the chemical potential
of the neutron becomes sufficiently large, energetic neutrons
can decay via weak strangeness non-conserving interactions
into $\Lambda$ hyperons leading to a $\Lambda$ Fermi sea
with $\mu_{\Lambda}=\mu_n$.
However, one expects $\Sigma^-$ to appear via
\begin{equation}
    e^-+n \rightarrow \Sigma^- +\nu_e,
\end{equation}
at lower densities than the $\Lambda$, even though $\Sigma^-$ is more
massive. The negatively charged hyperons
appear in the ground state of matter when their masses
equal $\mu_e+\mu_n$, while the neutral hyperon $\Lambda$
appears when its mass equals $\mu_n$. Since the
electron chemical potential in matter is larger than
the mass difference $m_{\Sigma^-}-m_{\Lambda}= 81.76$ MeV,
$\Sigma^-$ will appear at lower densities than $\Lambda$.
For matter with hyperons as well
the chemical equilibrium conditions become
\begin{eqnarray}
    \mu_{\Xi^-}=\mu_{\Sigma^-} = \mu_n + \mu_e, \nonumber \\
    \mu_{\Lambda} = \mu_{\Xi^0}=\mu_{\Sigma^0} = \mu_n , \nonumber \\
    \mu_{\Sigma^+} = \mu_p = \mu_n - \mu_e .
    \label{eq:beta_baryonicmatter}
\end{eqnarray}

Hyperonic degrees of freedom have been considered by several authors,
mainly within the framework of relativistic
mean field models \cite{prakash97,pke95,ms96} or parametrized
effective interactions \cite{bg97}, 
see also Balberg {\em et al.} \cite{blc99}
for a recent update. Realistic hyperon-nucleon interactions
were employed by Schulze {\em et al.} \cite{bbs98},
in a many-body calculation in order to study
the onset of hyperon formation in neutron star matter. In a recent
work \cite{bbs00}, they extend their work to study the properties
of neutron stars with hyperons, paying special attention to the
role played by three-body nucleon forces. 
All these works show that hyperons appear at densities of the order of
$\sim 2\rho_0$.

In Refs.~\cite{bbs98,bbs00} the hyperon-hyperon interaction was not 
included. However, it is clear that
as soon as the $\Sigma^-$ hyperon appears, one needs to consider the 
interaction between hyperon pairs
since  
it will influence the single-particle energy of hyperons, hence affecting
the equilibrium conditions from that density on and the density
where other hyperons (e.g. the $\Lambda$) appear. 
The aim of this work is thus 
to present results of 
many-body calculations for $\beta$-stable neutron star matter
with hyperonic degrees of freedom,
employing interactions
which also account for strangeness $S < -1$.
To achieve this goal,
our many-body scheme starts with the most recent
parametrization
of the free baryon-baryon potentials
for the complete  baryon octet
as defined by Stoks and Rijken in Ref.\
\cite{sr99}. 
This entails a microscopic
description of matter starting from
realistic nucleon-nucleon, hyperon-nucleon
and hyperon-hyperon interactions.
In a recent work \cite{isaac99}
we have developed a formalism for microscopic
Brueckner-type calculations of dense nuclear
matter that includes all types of
baryon-baryon interactions and allows to treat
any asymmetry on the fractions of the
different species ($n, p, \Lambda, \Sigma^-, \Sigma^0, \Sigma^+,
\Xi^-$ and $\Xi^0$).

Here we extend the calculations of Ref.\ \cite{isaac99}
to studies of $\beta$-stable neutron star matter.
A brief summary of the formalism
discussed in Ref.\ \cite{isaac99} is presented  
in section \ref{sec:sec2}. 
Our results are shown in section \ref{sec:sec3}. In
\ref{sec:sec3a} we discuss the
equation of state (EoS) and the composition of $\beta$-stable matter with
strangeness, using various nucleonic contributions to the EoS. 
Based on the composition of
matter we discuss in section \ref{sec:sec3b} the
possible neutron star structures. Our conclusions are given
in section \ref{sec:conclu}.

\section{Formalism}
\label{sec:sec2}

Our many-body scheme starts with the most recent
parametrization
of the free baryon-baryon potentials
for the complete  baryon octet
as defined by Stoks and Rijken in Ref.\
\cite{sr99}.
This potential model, which aims at describing all
interaction channels
with strangeness from $S=0$ to $S=-4$,
is based on SU(3) extensions
of the Nijmegen potential models \cite{rsy98}
for the $S=0$ and $S=-1$ channels, which
are fitted to the available body of experimental
data and constrain all free parameters in the model.
In our discussion we employ
the interaction version NSC97e of Ref.\ \cite{sr99}, since this
model, together with the model NSC97f of Ref.\ \cite{sr99}, results in
the best predictions for hypernuclear observables \cite{rsy98}.
For a discussion of other interaction models, see Refs.\ 
\cite{sr99,sl99}.

With a given interaction model,
the next step is to introduce effects from the nuclear medium.
Here we will construct the so-called $G$-matrix, which
takes into account short-range correlations for all strangeness
sectors, and solve the equations for the single-particle energies
of the various baryons self-consistently.
The $G$-matrix is formally given by
\begin{eqnarray}
   \left\langle B_1B_2\right |G(\omega)\left | B_3B_4 \right\rangle=
   \left\langle B_1B_2\right |V\left | B_3B_4 \right\rangle+&\nonumber\\
   \sum_{B_5B_6}\left\langle B_1B_2\right |V\left | B_5B_6 \right\rangle
   \frac{Q}{\omega-E_{B_5}-E_{B_6}+ \imath\eta}&\nonumber\\
   \times\left\langle B_5B_6\right |G(\omega)\left | B_3B_4 \right\rangle&.
   \label{eq:gmatrix}
\end{eqnarray}

\begin{figure}[hbtp]
 \setlength{\unitlength}{1mm}
       \begin{picture}(100,180)
       \put(15,10){\epsfxsize=12cm \epsfbox{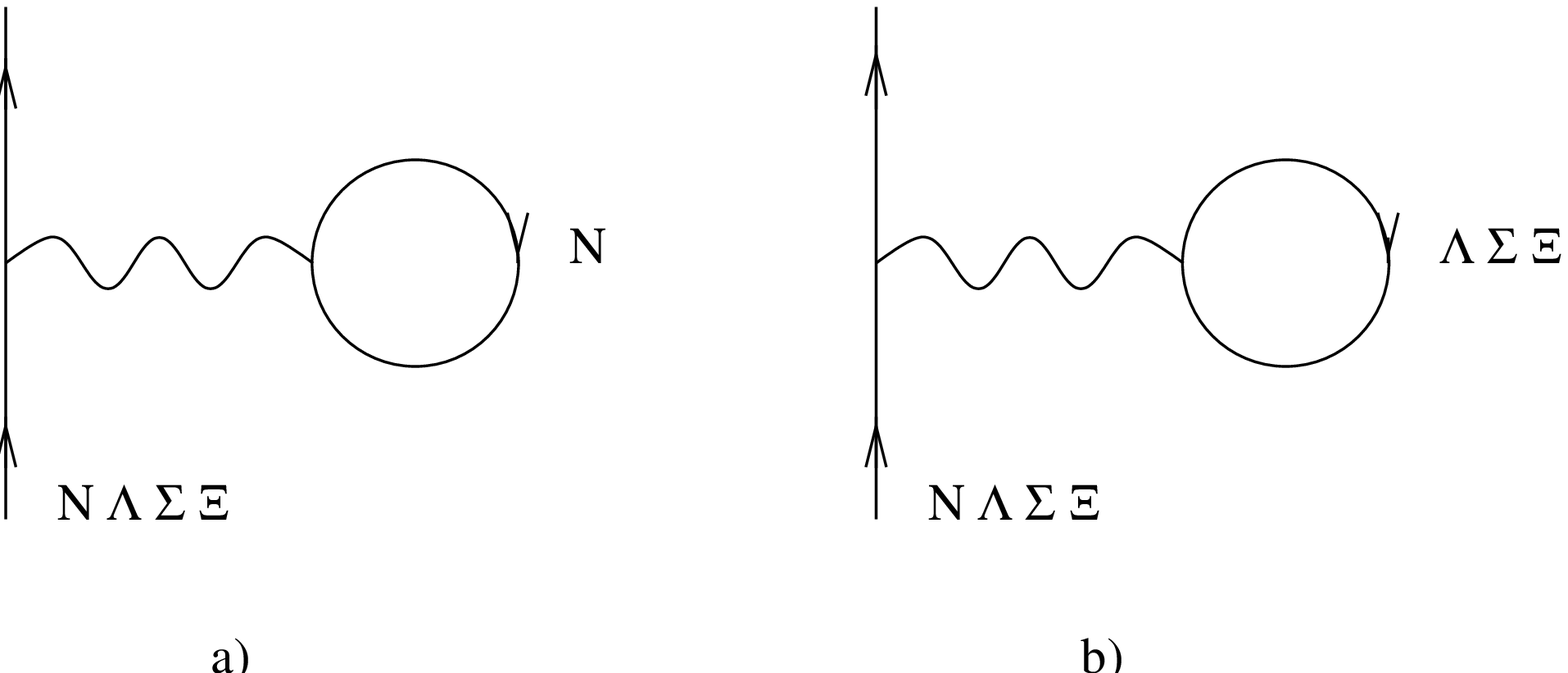}}
       \end{picture}
   \caption{Goldstone diagrams for the single-particle potential $U_B$.
            a) represents the contribution from nucleons only as hole
            states while b) includes only hyperons as hole states.
            The wavy line represents the $G$-matrix.}
   \label{fig:upot}
\end{figure}

Here $B_i$ represents all possible baryons $n$, $p$, $\Lambda$, $\Sigma^{-}$,
$\Sigma^0$, $\Sigma^+$, $\Xi^-$ and $\Xi^0$ and their quantum numbers
such as spin, isospin, strangeness, linear momenta and orbital momenta.
$Q$ is the Pauli operator which allows only intermediate states $B_5B_6$ 
compatible with the Pauli principle, and the energy variable $\omega$ is 
the starting 
energy
defined by the single-particle energies
of the incoming external particles $B_3B_4$.
The $G$-matrix is solved using relative and centre-of-mass coordinates,
see e.g., Refs.~\cite{isaac99,sl99} for computational details.
The single-particle energies are given by
\begin{equation}
      E_{B_i}=T_{B_i} + U_{B_i} +M_{B_i}
       \label{eq:spenergy}
\end{equation}
where $T_{B_i}$ is the kinetic energy and $M_{B_i}$
the mass of baryon ${B_i}$. 
We note that the $G$-matrix, Eq. (\ref{eq:gmatrix}), has been solved
using 
the standard prescription (i.e. $E_{B}=T_B+M_B$)
for the
intermediate states $B_5B_6$.
The single-particle potential $U_{B_i}$ is defined by 
\begin{equation}
       U_{B_i}={\mathrm Re} \sum_{B_j\leq F_j}
       \left\langle B_iB_j\right |
       G(\omega=E_{B_j}+E_{B_i})
       \left | B_iB_j \right\rangle \ ,
\label{eq:upot}
\end{equation}
where the linear momentum of the intermediate
single-particle state $B_j$ is limited by the size of the Fermi surface
$F_j$ for particle species $B_j$. The matrix element in Eq.
(\ref{eq:upot}) is
properly 
antisymmetrized
when the species $B_i$ and $B_j$ are the same.
This equation is displayed in terms of Goldstone diagrams
in Fig.\ \ref{fig:upot}. Diagram (a) represents contributions
from nucleons only as hole states, while diagram (b)
has only hyperons as holes states in case we have a finite hyperon
fraction in $\beta$-stable neutron star matter. The external legs
represent nucleons and hyperons. 
Detailed expressions for the single-particle energies and the $G$-matrices
involved can be found in Ref.\ \cite{isaac99}. 

The total non-relativistic energy density, $\varepsilon$, measured with
respect to the nucleon mass, is
obtained by adding the non-interacting 
leptonic contribution, $\varepsilon_l$, and  the
baryonic contribution, $\varepsilon_b$, the latter being obtained
from the baryon
single-particle potentials
\begin{equation}
\varepsilon=\varepsilon_l+\varepsilon_b=\varepsilon_l+2\sum_{B}
\int_0^{k_F^{(B)}} \frac{d^3 k}{(2\pi)^3}
\left(M_B + \frac{\hbar^2k^2}{2M_B}+\frac{1}{2}U_B(k) - M_N \right) \ .
\label{eq:binding}
\end{equation}
The total binding
energy per baryon, ${\cal E}$, is then given by
\begin{equation}
{\cal E}=\frac{\varepsilon}{\rho} \ ,
\label{eq:binding2}
\end{equation}
where $\rho$ is the total baryonic density. The density of a given fermion 
species is given by
\begin{equation}
\rho_{i}=\frac{k_{F_i}^3}{3\pi^2}=x_{i}\rho_T \ ,
\end{equation}
where
$x_i$ is the fraction of particle species $i$ and
$\rho_T=\rho+\rho_l$ is the total density which includes the
baryonic ($\rho$) and 
the leptonic one ($\rho_l$).

In order to satisfy the equations 
for $\beta$-stable matter summarized in Eq.\ (\ref{eq:beta_baryonicmatter}), 
we need to know the chemical potentials of the particles involved.
In Brueckner-Hartree-Fock (BHF) theory the chemical potential is taken as
the
single particle energy at the Fermi momentum of the baryon, $k_F^{(B)}$,
which at the lowest order reads
\begin{equation}
\mu_B=E_B(k_F^{(B)})=M_B+T_B(k_F^{(B)})+U_B^N(k_F^{(B)})+
U_B^Y(k_F^{(B)}) \ ,
\label{eq:chempot}
\end{equation}
where, in the last equality, the baryon
single-particle potential $U_B$ has been split into a contribution,
$U_B^N$, coming from the nucleonic Fermi seas ($p, n$) and a contribution,
$U_B^Y$,
coming from the hyperonic ones ($\Sigma^-, \Lambda, \dots$).
From calculations in pure nucleonic matter it is well known
that the nucleon chemical potential obtained from of 
Eq. (\ref{eq:chempot}) differs considerably
from the thermodynamic definition
\begin{equation}
\mu_B= \frac{\partial \varepsilon}{\partial \rho_B}  \ .
\end{equation}
Therefore, for the nucleons, we replace the nucleonic contribution 
to the
chemical potential in Eq. (\ref{eq:chempot}), i.e. $\mu_N^N=
M_N+T_N(k_F^{(N)})+U_N^N(k_F^{(N)})$,
by $\mu_N^N=\partial \varepsilon_{NN}/\partial \rho_N$, 
where 
\begin{equation}
\varepsilon_{NN}=
2\sum_{B=n,p}
\int_0^{k_F^{(B)}} \frac{d^3 k}{(2\pi)^3}
\left(M_B + \frac{\hbar^2k^2}{2M_B}+\frac{1}{2}U_B^N(k) \right) 
\label{eq:nnepsilon}
\end{equation}
is the nucleonic contribution to the baryonic
energy density including only the interaction between $NN$ pairs.
For the hyperons, we keep the prescription of Eq. (\ref{eq:chempot}).
As shown in Ref. \cite{bbs00}, these approximations amount to ignoring
the weak dependence of $U_N^N, U_Y^N$ on the hyperon fractions and
of $U_N^Y, U_Y^Y$ on the nucleon ones, and are good enough as long
as the proton and hyperon fractions keep moderately small. 
Using the parabolic approximation for 
$\varepsilon_{NN}$
one obtains
\cite{bbs00}
\begin{equation}
\mu_{p,n}(\rho_N,\beta)=\mu_{p,n}(\rho_N,\beta=0)-\left(\beta^2 \pm 2\beta
-\beta^2\rho_N
\frac{\partial}{\partial \rho_N}\right)E_{sym}(\rho_N) \ ,
\end{equation}
where $\beta=1-2\rho_p/\rho_N$ is the asymmetry parameter, with
$\rho_N=\rho_n+\rho_p$. The symmetry energy
$E_{sym}$
can be expressed as the difference of the energy per nucleon, $\tilde{\cal E}$,
between pure neutron
($\beta=1$) and symmetric nuclear ($\beta=0$) matter:
\begin{equation}
E_{sym}(\rho_N)=\tilde{\cal E}(\rho_N,\beta=1)-\tilde{\cal
E}(\rho_N,\beta=0)=\frac{1}{2}\frac{\partial
\tilde{\cal
E}}{\partial \beta}(\rho_N, \beta=1) \ ,
\end{equation}
where $\tilde{\cal E}$ is the nucleonic contribution to the total
energy per nucleon, $ \tilde{\cal E} = \varepsilon_{NN}/\rho_N$, and
$\mu_{p,n}(\rho_N,\beta=0)$ is given by
\begin{equation}
\mu_{p,n}(\rho_N,\beta=0)=\tilde{\cal E}(\rho_N,\beta=0)+
\rho_N\frac{\partial \tilde{\cal
E}(\rho_N,\beta=0)}{\partial \rho_N}.
\end{equation}


The many-body approach outlined above is the lowest-order
BHF method extended to the hyperon sector.
This means also that we consider only
two-body interactions. However, it is well-known from studies of nuclear
matter and neutron star matter with nucleonic degrees of freedom only
that three-body forces are important in 
order to reproduce the saturation 
properties of nuclear matter, see e.g., Ref.\ \cite{apr98} for the most 
recent
approach.  
The effect of nucleon three-body forces on the properties of
$\beta$-stable matter with hyperons has been studied in 
Refs. \cite{bbs98,bbs00}. It is found that the repulsion induced
by the three-body force at high densities enhances
substantially the hyperon population and produces a strong
softening of the EoS.
In order to include such effects, we will alternatively use for the nucleonic sector,
the EoS of Ref.\ \cite{apr98} (hereafter referred to as APR98), which is
obtained from a variational calculation using the Argonne $V_{18}$
nucleon-nucleon interaction \cite{v18} with relativistic boost corrections
and a fitted three-body interaction model. 

In the discussions below we will thus present two sets of results for
$\beta$-stable matter, one where the nucleonic contributions
to the self-energy of nucleons are derived from the baryon-baryon potential
model of Stoks  and Rijken \cite{sr99} and one where the nucleonic 
contributions to the neutron and proton chemical potentials are
calculated from the parametrization of the APR98 EoS discussed in 
Eq. (49) of  Ref.\ \cite{hh99}. We note that replacing the nucleon-nucleon 
part of the interaction model of
Ref.\ \cite{sr99} with that from the
$V_{18}$ nucleon-nucleon interaction \cite{v18}, 
does not introduce large differences at the BHF level. However,
the inclusion of three-body forces as done in Ref.\ \cite{apr98} is 
important.
Hyperonic contributions
will however all be calculated with the baryon-baryon interaction of
Stoks  and Rijken \cite{sr99}. We emphasize that, in the present work, 
$YN$ as well as $YY$ interactions are taken into account.

\section{Results}
\label{sec:sec3}

\subsection{Equation of state and composition of $\beta$-stable matter}
\label{sec:sec3a}

\begin{figure}[hbtp]
 \setlength{\unitlength}{1mm}
       \begin{picture}(100,180)
       \put(15,10){\epsfxsize=12cm \epsfbox{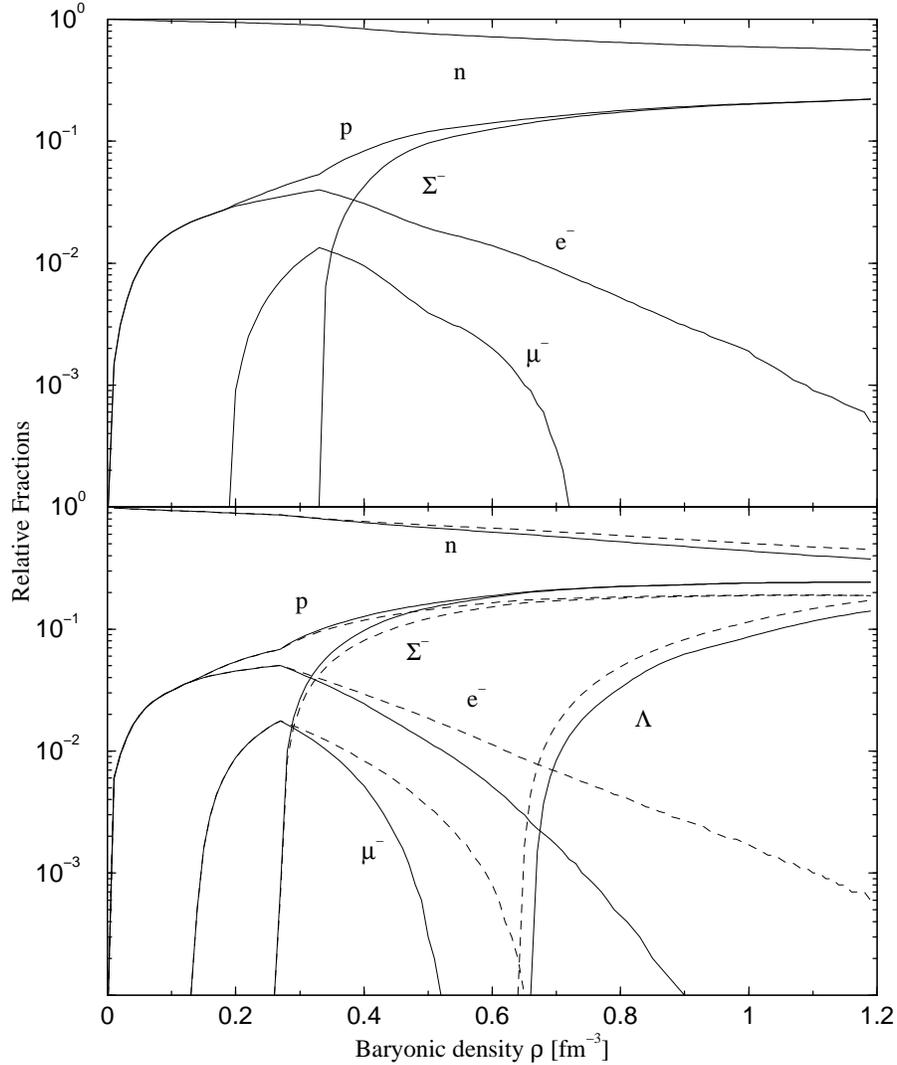}}
       \end{picture}
  
   \caption{Composition of $\beta$-stable neutron star matter. In the
            upper panel results for the Stoks and Rijken potential 
            \protect\cite{sr99} are presented. 
            In the  lower panel the nucleonic part of the self-energy of 
            the nucleons has been replaced with the EoS of Ref.\ 
            \protect\cite{apr98}. Solid lines in upper and lower
            panel correspond to the case in which all the interactions
            (nucleon-nucleon, hyperon-nucleon and hyperon-hyperon) are
            considered. Dashed lines in the lower panel correspond
            to the case where the hyperon-hyperon interaction has been 
            switched off.}
   \label{fig:fraction}
\end{figure}

The above models for the pure nucleonic part (NSC97e and APR98)
combined with the hyperon contribution (NSC97e)
yield the composition of $\beta$-stable matter, up to total
baryonic density $\rho=1.2$ fm$^{-3}$, shown in Fig.\  \ref{fig:fraction}.
In the upper panel, results for NSC97e model for $NN$, $YN$ and $YY$
interactions are presented. Results combining the APR98 model for
the nucleonic sector with the 
NSC97e for the $YN$ and $YY$ interactions are shown in the lower panel. 
In both
panels, solid lines correspond to the case in which all the interactions
$NN$, $YN$ and $YY$ are considered. 

\begin{figure}
 \setlength{\unitlength}{1mm}
       \begin{picture}(100,180)
       \put(15,10){\epsfxsize=12cm \epsfbox{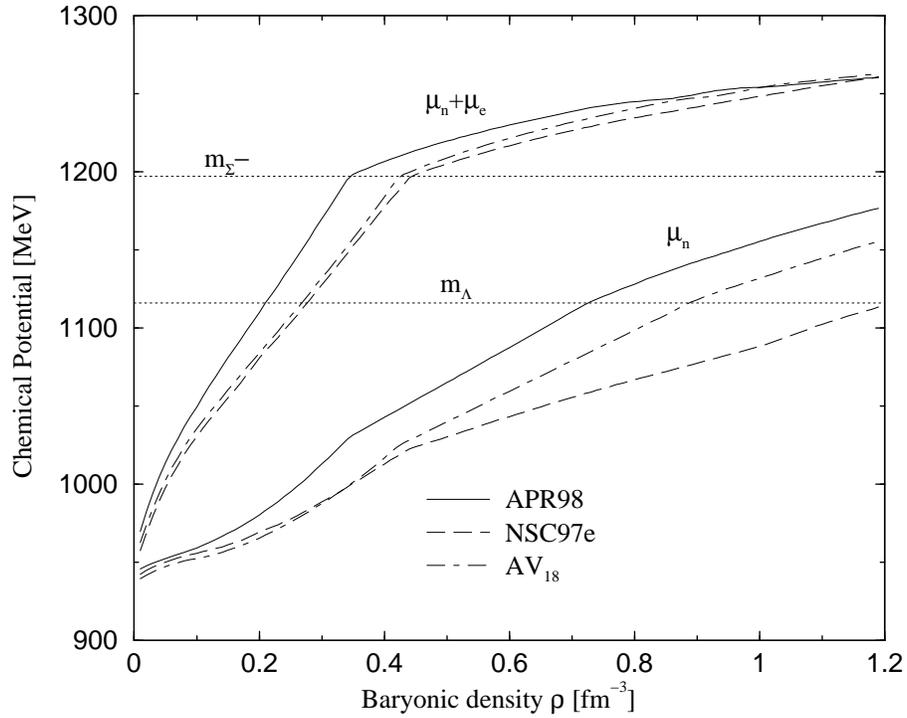}}
       \end{picture}
   \caption{Chemical equilibrium condition for the appearence of $\Sigma^-$ and
            $\Lambda$ hyperons for the case of free hyperons and
            three different nucleon-nucleon interactions. Dotted
            straight lines denote the rest masses of the hyperons.}
   \label{fig:freechempot}
\end{figure}

As can be seen by comparing the solid lines in both
panels in Fig.\  \ref{fig:fraction},
the composition of $\beta$-stable neutron star matter has a strong
dependence on the model used to describe the non-strange sector. 
In both cases, $\Sigma^-$ is
the first hyperon to appear due to its negative charge. Since the 
APR98 EoS yields a stiffer pure nucleonic matter EoS than the
corresponding one for NSC97e, the onset of $\Sigma^-$ for
the APR98 case occurs
at a smaller density ($\rho=0.27$
fm$^{-3}$) than for the NSC97e case ($\rho=0.34$ fm$^{-3}$). In
both cases, as soon as the $\Sigma^-$ hyperon appears, 
leptons tend to disappear,
totally in the APR98 case (the electron chemical potential changes
sign at $\rho=1.01$
fm$^{-3}$, signaling the appearance of positrons), whereas in the NSC97e case only muons disappear. 
The onset of
$\Lambda$ 
formation takes place at higher density.
Recalling the condition for the appearance of $\Lambda$,
$\mu_{\Lambda}=\mu_{n}=\mu_{p}+\mu_{e^-}$, and that the APR98 EoS is
stiffer due to the inclusion of three-body forces, this clearly enhances
the possibility of creating $\Lambda$ hyperons at lower
density with this interaction model
with respect to the NSC97e case. Indeed,
the APR98 model produces $\Lambda$ hyperons from
$\rho=0.67$ fm$^{-3}$ on, whereas the neutron chemical
potential of the NSC97e model turns out to be too small to equal
the $\Lambda$ one in the range of densities explored.   
The absence of $\Lambda$ hyperons in the NSC97e case
can also, in addition to a softer EoS, be retraced to a delicate
balance between the nucleonic and hyperonic hole state contributions
(and thereby to features of the baryon-baryon interaction)
to the self-energy of the baryons considered here, see diagrams (a) and (b)
in Fig.\ \ref{fig:upot}. Stated differently, the contributions from 
$\Sigma^-$, proton and neutron hole states to the $\Lambda$ chemical
potential are not attractive enough to lower the chemical potential of
the $\Lambda$ so that it equals that of the neutron. Furthermore,
the increase of the chemical potential of the neutron with
density is slowed down with the NSC97e $YN$ interaction model since
contributions from $\Sigma^-$ hole states to the neutron self-energy are
attractive, see e.g., Ref.\ \cite{isaac99} for a detailed account of these
aspects of the interaction model. 
We note that the 
isospin-dependent component (Lane term) of the        
$\Sigma^-$ single-particle potential for the
new Nijmegen interaction \cite{sr99} is strongly attractive,
as opposed to what is found \cite{dabro99} for other interactions,
including the old Nijmegen one \cite{nij89}. This 
in turn implies a strong
attraction for $\Sigma^- n$ ($T=3/2$) pairs, which is 10 times that
obtained for the old Nijmegen potential at saturation density. These differences
become more noticeable as density increases: 
while the $\Sigma^- n$ pairs become increasingly
more attractive with the new Nijmegen potentials (see e.g. Fig. 5
in \cite{isaac99}), they turn out to be strongly repulsive for the
old one (see e.g. Figs. 1 and 2 in \cite{bbs00}). This is why in Ref.\ 
\cite{bbs00} the onset density for the appearance of $\Sigma^-$ is larger
than that for free hyperons, whereas the reverse situation is found here
(compare the $\Sigma^-$ onset point in Fig.\  \ref{fig:fraction} with what
would be extracted from Fig.\ \ref{fig:freechempot}).
Within our many-body approach, no other
hyperons appear at densities below $\rho=1.2$ fm$^{-3}$. 
These results differ from present mean field calculations
\cite{prakash97,pke95,ms96}, where all kinds of hyperons can appear at
the densities considered here.
Although the APR98 EoS may be viewed as the currently
most realistic approach to the nucleonic EoS, our results have to be
gauged with the uncertainty in the hyperon-hyperon and hyperon-nucleon
interactions. Especially, if the hyperon-hyperon interactions tend to be
more attractive, this may lead to the formation of hyperons such as the
$\Lambda$, $\Sigma^0$, $\Sigma^+$, $\Xi^-$ and $\Xi^0$ at lower
densities. The stiffness of the nucleonic contribution, together with the 
hyperon-nucleon and hyperon-hyperon interactions 
play crucial roles in the appearance of various
hyperons beyond the $\Sigma^-$. 

In order to examine the role of the hyperon-hyperon interaction on the
composition of $\beta$-stable neutron star matter, the dashed lines in the
lower panel of Fig.\  \ref{fig:fraction} show the results of a calculation 
in which only the $NN$ and $YN$ interactions are taken into account. When
the $YY$ interaction is switched off, the scenario described above changes
only quantitatively. The onset point of $\Sigma^-$ does not change,
because $\Sigma^-$ is the first hyperon to appear and therefore the $YY$
interaction plays no role for densities below this point. 
We note that
the reduction of the $\Sigma^-$ fraction compared with the 
case which includes the $YY$ interaction, is a consequence of neglecting
the strongly attractive $\Sigma^-\Sigma^-$ interaction \cite{sr99}, which
allows the energy balance ($n n \leftrightarrow \Sigma^- p$) to be 
fulfilled with a smaller
$\Sigma^-$ Fermi sea.
In turn, the reduction of the $\Sigma^-$ fraction 
yields a moderate 
increase of the leptonic content in order to keep charge neutrality (in
fact only muons disappear now). On the other hand, a smaller amount of $\Sigma^-$'s
implies less
$\Sigma^-$$n$ pairs. Recalling that the $\Sigma^-$$n$
interaction is attractive in this model 
(see e.g. Fig.\ 7 of Ref.\ \cite{isaac99}),
this means that the chemical potential of the neutrons becomes now less
attractive. As a consequence, the $\Lambda$ hyperons appear at a smaller
density ($\rho=0.65$ fm$^{-3}$)
and have a larger relative fraction.      

As it has been mentioned, the composition of $\beta$-stable matter 
depends strongly on the model used to describe the nucleonic sector. 
In order to study this dependence, Fig.\
\ref{fig:freechempot} shows the chemical potential of the neutron and 
the sum
$\mu_{n}+\mu_{e^-}$ for 
$\beta$-stable matter composed of nucleons and 
free hyperons for three different $NN$ interaction models.
The solid lines correspond to the variational calculation denoted
by APR98, which uses the Argonne $V_{18}$ interaction and includes
three-body forces and relativistic boost corrections. The dashed
and dot-dashed lines correspond to lowest-order BHF calculations
using, respectively, the NSC97e and the Argonne $V_{18}$ potentials, the
latter extracted from the results of Ref. \cite{bbs00}.
Dotted lines denote the $\Lambda$ and
$\Sigma^-$ masses. In this case, the onset conditions for $\Lambda$ and
$\Sigma^-$ are, respectively, $\mu_{n}=m_{\Lambda}$ and
$\mu_{n}+\mu_{e^-}=m_{\Sigma^-}$. As can bee seen from the
figure, the onset points of both hyperons are
different depending on the $NN$ interaction model employed. 
In the APR98 model both the $\Sigma^-$ and $\Lambda$ hyperons appear 
at lower densities than in the lowest-order BHF models using the Argonne
$V_{18}$ or the NSC97e interactions. 
This is a consequence of the
different stiffness of the EoS generated by the three $NN$ interaction
models. 
In fact, the NSC97e interaction gives the
softest EoS and it is not even able to produce $\Lambda$ hyperons in
the range of densities explored. 
Note that the hyperon onset points determined from
Fig.~\ref{fig:freechempot}
differ slightly from those observed in Fig.~\ref{fig:fraction} 
as a consequence of the effect of the $YN$ and
$YY$ interactions. Since the differences are not so large one
concludes that
the main features of the composition of matter are dominated by
the pure nucleonic contribution to the EoS.

At present, the APR98 EoS represents perhaps
the most sophisticated many-body approach to nuclear matter. Therefore,
in what follows, we restrict our results to this $NN$ interaction model 
supplemented with the NSC97e one 
for the hyperonic sector. 

\begin{figure}[hbtp]
 \setlength{\unitlength}{1mm}
       \begin{picture}(100,180)
       \put(15,10){\epsfxsize=12cm \epsfbox{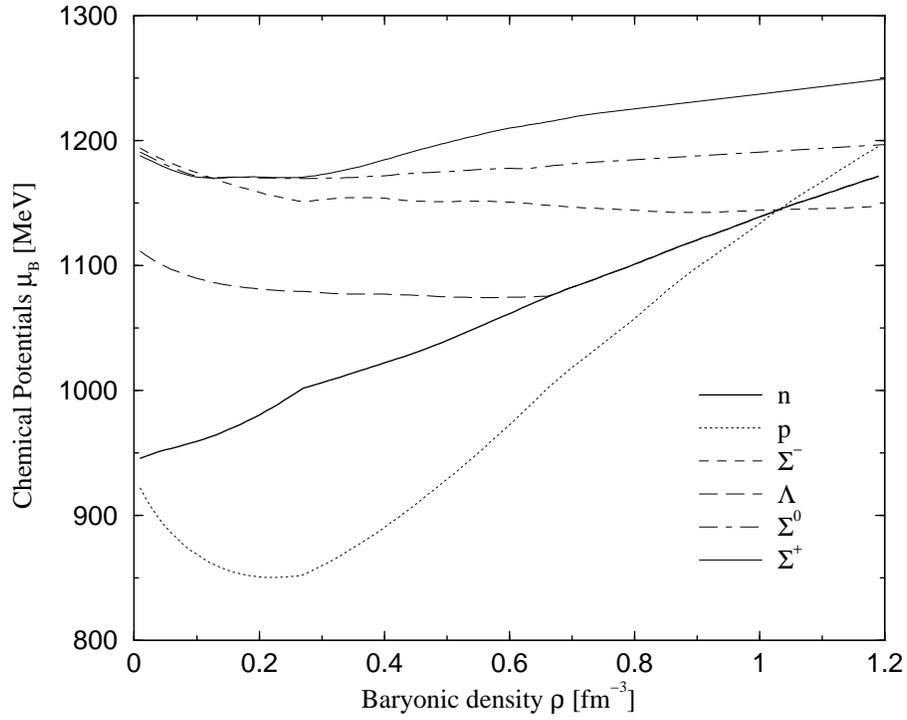}}
       \end{picture}
   \caption{Chemical potentials in $\beta$-stable neutron star matter
            as functions of the total baryonic density $\rho$.}
   \label{fig:chempots}
\end{figure}

In Fig.\ \ref{fig:chempots} we show
the chemical potentials in $\beta$-stable neutron star matter for
different baryons. We note that  
neither the $\Sigma^0$ nor the $\Sigma^+$
do appear since, as seen from the figure, the respective stability criteria
of Eq. (\ref{eq:beta_baryonicmatter}) are not fulfilled.
This is due, partly, to the fact that none of the $\Sigma^0$-baryon 
and the
$\Sigma^+$-baryon self-energies
are attractive enough. A similar argument applies to
$\Xi^0$ and $\Xi^-$. In the latter case the mass of the particle is $\sim
1315$ MeV and an attraction of around $200$ MeV would be needed to fulfill the
condition $\mu_{\Lambda}=\mu_{\Xi^0}=\mu_{n}$ at the
highest density explored in this work. From the figure we see,
however, that the $\Sigma^0$ hyperon could appear at densities 
close to $1.3$
fm$^{-3}$. 

Fig.\ \ref{fig:eosfig} shows the EoS for four different cases:
pure nucleonic matter (solid line); matter with nucleons and
free hyperons (dotted line); matter with nucleons and hyperons
interacting only with nucleons (dashed line) and, finally, matter with nucleons and
hyperons interacting both with nucleons and hyperons (long-dashed line). Each
curve corresponds to a different composition of $\beta$-stable neutron
star matter, obtained by requiring the equilibrium conditions of
Eq. (\ref{eq:beta_baryonicmatter}), with the appropriate
chemical potentials for each of the four cases. 
The leptonic contribution to the EoS is also included in all the
cases. As can be seen comparing the solid and dotted lines the
appearance of hyperons leads to a considerable softening of the EoS.
This softening is essentially due to a decrease of the kinetic energy
because the hyperons can be accomodated in lower momentum
states and in addition have a large bare mass. The 
hyperon-nucleon interaction (dashed line) has two effects. On
one hand, for densities
up to $\rho \sim 0.72$ fm$^{-3}$, the $YN$
interaction reduces the total energy per baryon making
the EoS even softer. On the other hand, for 
densities higher than $\rho= 0.72$ fm$^{-3}$, it becomes repulsive and
therefore, the EoS becomes slightly stiffer than that for  
non-interacting hyperons. The contribution from
the hyperon-hyperon interaction (long-dashed line) is always attractive
producing a softening of the EoS over the whole range of densities explored.
We note that, at high densities, the combined
incorporation of the repulsive $YN$ and attractive $YY$ interactions 
causes the
EoS to become closer to that for 
non-interacting hyperons.

\subsection{Structure of neutron stars}
\label{sec:sec3b}

We end this section with a discussion on neutron star properties 
with the above equations of state.

The best determined neutron star masses are found in binary pulsars
and all lie in the range $1.35\pm 0.04 M_\odot$ \cite{tc99}
except for the nonrelativistic pulsar PSR J1012+5307
of mass
$M=(2.1\pm 0.8)M_\odot$ \cite{vanParadijs1998}. Several X-ray binary
masses have been measured of which the heaviest are
Vela X-1 with $M=(1.9\pm 0.2)M_\odot$ \cite{Barziv1999}
and Cygnus X-2 with
$M=(1.78\pm 0.2)M_\odot$ \cite{OroszKuulkers1999}.  The recent
discovery of high-frequency brightness oscillations in low-mass X-ray
binaries provides a promising new method for determining masses and
radii of neutron stars, see Ref.\ \cite{miller99}. The
kilohertz quasi-periodic oscillations (QPO) occur in pairs
and are most likely the orbital
frequencies of accreting matter
in Keplerian orbits around neutron stars of mass $M$ and its beat
frequency with the neutron star spin.  According to
Zhang {\em et al.} \cite{zhang98} and   Kaaret {\em et al.}
\cite{kaaret1998} the accretion
can for a few QPO's be tracked to its innermost stable orbit.
For slowly rotating stars the resulting mass is
$M\simeq2.2M_\odot({\mathrm{kHz}}/\nu_{QPO})$.  For example, the
maximum frequency of 1060 Hz upper QPO observed in 4U 1820-30 gives 
$M\simeq 2.25M_\odot$ after correcting for the neutron star
rotation frequency.  If the maximum QPO frequencies of 4U 1608-52
($\nu_{QPO}=1125$ Hz) and 4U 1636-536 ($\nu_{QPO}=1228$ Hz) also
correspond to innermost stable orbits the corresponding masses are
$2.1M_\odot$ and $1.9M_\odot$. 
These observations define a range of variation for the mass between
$M\sim 1.35 M_\odot$ and $M\sim 2.2 M_\odot$, which severely restricts the
EoS for dense matter. 

In order to obtain the radius and mass of a neutron star, we have solved
the Tolman-Oppenheimer-Volkov equation with and without
rotational corrections, following the approach of Hartle \cite{hartle1967},
see also Ref.\ \cite{hh99}.

\begin{figure}
 \setlength{\unitlength}{1mm}
       \begin{picture}(100,180)
       \put(15,10){\epsfxsize=12cm \epsfbox{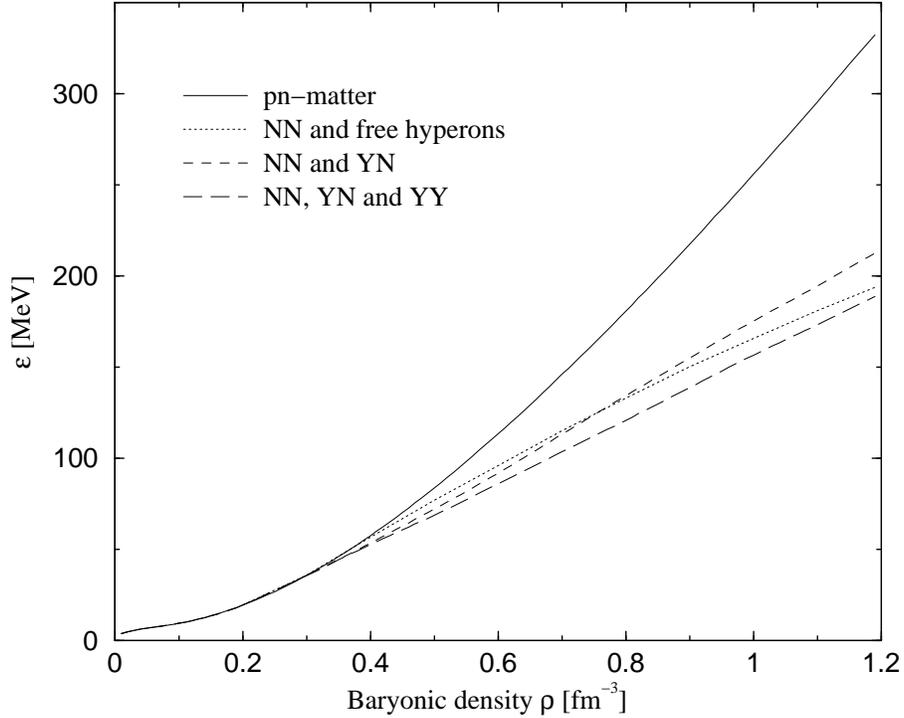}}
       \end{picture}
   \caption{Energy per baryon in $\beta$-stable neutron star matter 
            as function of the total baryonic density $\rho$, for four 
            cases: pure nucleonic matter (solid lines); matter with
            nucleons and non-interacting hyperons (dotted line); matter
            with nucleons and hyperons interacting only with nucleons
            (dashed line); and matter with nucleons and hyperons
            interacting with nucleons and hyperons (long-dashed line).
            The leptonic contribution to the energy per baryon is
            included in all cases.}
    \label{fig:eosfig}
\end{figure}

Following the discussion of Figs.~\ref{fig:freechempot}, 
\ref{fig:chempots} and \ref{fig:eosfig}, we use the 
pure nucleonic matter EoS of Akmal {\em et al.} \cite{apr98},
including both relativistic boost corrections and
three-body interactions. 
Relating to the discussion of the previous subsection, we study
two additional cases as well. One case 
where we just include the hyperon-nucleon
interaction as done in e.g., Refs.~\cite{bbs00} and another case
where we include both the hyperon-nucleon and the hyperon-hyperon
interactions.

\begin{figure}\begin{center}
 \setlength{\unitlength}{1mm}
       \begin{picture}(100,180)
       \put(-10,10){\epsfxsize=12cm \epsfbox{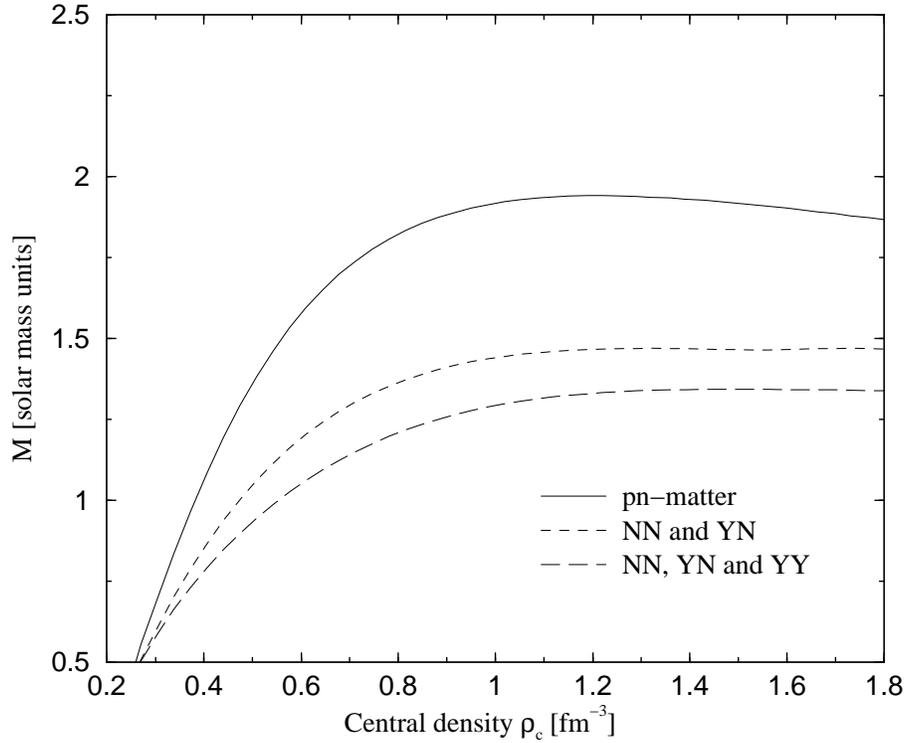}}
       \end{picture}
   \caption{Total mass $M$ for various equations of state. Rotational
            corrections are not included. The solid line corresponds to the 
            case of $\beta$-stable matter with nucleonic degrees of freedom 
            only, 
            the short-dashed line includes also the effects of the 
            hyperon-nucleon interaction, while the long-dashed line includes the
            three types of baryon-baryon interactions.}
   \label{fig:mass}
\end{center}\end{figure}

\begin{figure}\begin{center}
 \setlength{\unitlength}{1mm}
       \begin{picture}(100,180)
       \put(-10,10){\epsfxsize=12cm \epsfbox{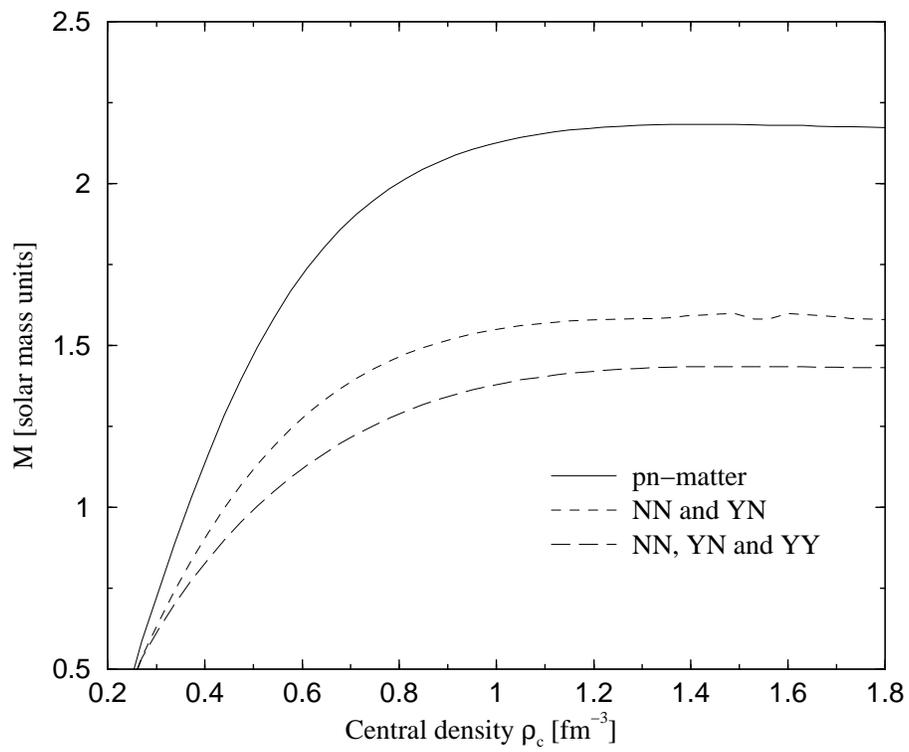}}
       \end{picture}
   \caption{Same as previous figure, but with rotational corrections.}
   \label{fig:mass1}
\end{center}\end{figure}

\begin{figure}\begin{center}
 \setlength{\unitlength}{1mm}
       \begin{picture}(100,180)
       \put(-10,10){\epsfxsize=12cm \epsfbox{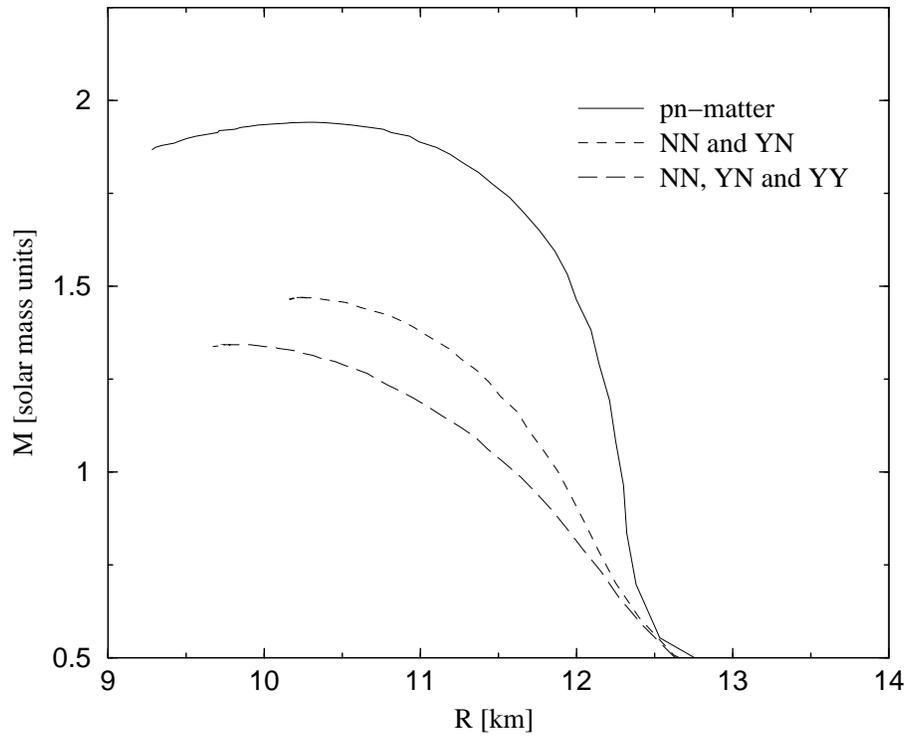}}
       \end{picture}
   \caption{Mass-radius relation without rotational corrections.
            Notations as in the previous two figures.}
   \label{fig:massradius}
\end{center}\end{figure}

In Fig.\ \ref{fig:mass} we show the resulting mass as function
of these equations of state without rotational 
corrections. In Fig.~\ref{fig:mass1} 
we include rotational corrections to the mass. The mass-radius relations
for the same equations of state are shown in
Fig.~\ref{fig:massradius}.

For $\beta$-stable pure nucleonic matter 
the EoS is rather stiff compared
with the EoS obtained with hyperons, see Fig.\ \ref{fig:eosfig}.
The EoS yields a maximum mass $M= 1.89M_\odot$ without
rotational corrections and $M=2.11M_\odot$ when rotational
corrections are included. The results for the masses as functions of central
density $\rho_c$ are shown by the solid lines 
in Figs.\ \ref{fig:mass} and \ref{fig:mass1}.
The corresponding mass-radius relation for masses without rotational
corrections are shown in Fig.~\ref{fig:massradius}.

The second set of results refers to the case where we allow for the presence of 
hyperons and consider the 
hyperon-nucleon interaction but explicitly exclude the hyperon-hyperon one. The
corresponding results are shown by the short-dashed lines in Figs.~\ref{fig:mass},
\ref{fig:mass1} and \ref{fig:massradius}. 
Without rotational corrections, we obtain
a maximum mass $M=1.47M_\odot$ whereas the rotational correction 
increases the mass to  $M= 1.60 M_\odot$. 
Thus, the inclusion of the the $YN$ interaction with the corresponding 
formation of $\Sigma^-$ and $\Lambda$ leads to a reduction 
of the mass by $\sim 0.4-0.5 M_{\odot}$. 
This large reduction is mainly a consequence of the strong softening of the EoS due to 
appearance of hyperons.

The last EoS employed is that which combines the nucleonic
part of Ref.\ \cite{apr98} with the computed hyperon 
contribution including both the hyperon-nucleon and 
the hyperon-hyperon interactions. 
These results are shown by the long-dashed lines in Figs.~\ref{fig:mass},
\ref{fig:mass1} and \ref{fig:massradius}.
The inclusion of the hyperon-hyperon interaction leads
to a further softening of the EoS in Fig.~\ref{fig:eosfig}, and,
as can be seen from Fig.\ \ref{fig:mass}, 
this leads to an 
additional 
reduction of the total mass. 
Without rotational corrections we obtain
a maximum mass $M\sim 1.34M_\odot$, whilst the rotational correction 
increases the mass to  $M\sim 1.44 M_\odot$. The size of the 
reduction due to the presence of hyperons, $\Delta M\sim 0.6-0.7M_\odot$, and the obtained 
neutron star masses are comparable
to those reported by Balberg {\em et al.} \cite{blc99}.

There are other features as well to be noted from Figs.\ \ref{fig:mass},
\ref{fig:mass1} and \ref{fig:massradius}.
The EoS with hyperons reaches a maximum mass at a central density 
$\rho_c\sim 1.3-1.4$ fm$^{-3}$. In Fig.\ \ref{fig:fraction} 
we showed that the only hyperons which can appear at these densities
are $\Lambda$ and $\Sigma^-$.
If other hyperons were to appear at higher
densities, this would most likely lead to a further softening of the EoS, and 
thereby smaller neutron star masses. 

The reader should however note that our calculation of hyperon
degrees freedom is based on a non-relativistic BHF 
approach. Although the nucleonic part extracted from Ref.\ \cite{apr98},
including three-body forces and relativistic boost corrections, is to be 
considered as a benchmark calculation for nucleonic degrees of freedom,
relativistic effects in the hyperonic calculation could result in a
stiffer EoS and thereby larger mass. However, relativistic mean 
field calculations
with parameters which result in a similar composition of matter as
shown in Fig.\ \ref{fig:fraction}, result in similar masses
as  those reported in Fig.\ \ref{fig:mass}.
In this sense, our results with and without hyperons 
may provide a lower and upper bound for the 
maximum mass. This  leaves two natural options
when compared to the observed neutron star masses.
If the above heavy neutron stars prove erroneous 
by more detailed observations
and only masses like those of binary pulsars are found, 
this may indicate that heavier neutron stars simply are not
stable which in turn implies a soft EoS, or that a
significant phase transition must occur already at a few times nuclear
saturation densities. Our EoS with hyperons would fit into this case, although
the mass without rotational corrections is on the lower side.
Else, if the large masses from QPO's are confirmed, then the EoS
for baryonic matter needs to be stiffer. This would then pose a severe
problem to present hadronic models since, when the nucleonic part of the
EoS is 
sufficiently stiff to support large masses, one cannot avoid the
appearance of hyperons which, in turn, produce a softening of the EoS.

Although we have only considered the formation of hyperons in neutron
stars, transitions to other degrees of freedom such as quark matter,
kaon condensation and pion condensation may or may not take place
in neutron star matter.
We would however like to emphasize that the hyperon formation mechanism
is perhaps the most robust one and is likely to occur in the interior
of a neutron star, unless the hyperon self-energies are strongly repulsive 
due
to repulsive hyperon-nucleon and hyperon-hyperon interactions, a repulsion
which  would contradict
present data on hypernuclei \cite{bando}.
The EoS with hyperons yields however neutron star masses without
rotational corrections which are even below $\sim 1.4M_\odot$.
This means that our EoS with hyperons needs to be stiffer, 
a fact which may in turn
imply that more complicated many-body terms not included in our 
calculations,
such as three-body forces between nucleons and hyperons and/or relativistic
effects,  are needed.

\section{Conclusions}
\label{sec:conclu}

Employing the recent 
parametrization
of the free baryon-baryon potentials
for the complete  baryon octet
of Stoks and Rijken \cite{sr99}, we have performed a
microscopic many-body calculation of the structure of
$\beta$-stable neutron star matter including 
hyperonic degrees of freedom. 
The potential model employed allows for the presence of only two
types of hyperons up to densities of about ten times nuclear matter
saturation density. These hyperons are
$\Sigma^-$ and $\Lambda$. The interactions for strangeness
$S=-1$, $S=-2$, $S=-3$ and $S=-4$ are not attractive enough to allow
the formation of other hyperons. 

The presence of hyperons leads however
to a considerable softening of the EoS, entailing a corresponding reduction
of the maximum mass of the neutron star. With hyperons, including only
the hyperon-nucleon interaction, we obtain maximum
masses of the order $M=1.47M_\odot$ and $M=1.60 M_\odot$ without and 
with rotational corrections, respectively.
The inclusion of the hyperon-hyperon interaction leads to a
further softening of the EoS and
reduces the obtained masses to 
$M=1.34M_\odot$ and $M=1.44 M_\odot$ without and 
with rotational corrections, respectively.

The corresponding numbers with just nucleonic degrees of freedom are
$M=1.89 M_\odot$ and $M=2.11 M_\odot$, showing that the 
reduction in mass due to hyperonic degrees of freedom is 
$\sim 0.5-0.7M_\odot$.

Our novel result is that a further softening of the EoS is obtained when 
including the effect of the $YY$ interaction since it is attractive over 
the whole density range explored. This is mainly due to the $\Sigma\Sigma$
interaction which is strongly enough to develop a bound state \cite{sr99}.
We note that the $\Lambda\Lambda$ attraction produced by this model is only
mild, not 
being able to reproduce the experimental $2$$\Lambda$ separation energy of
$\Delta B_{\Lambda\Lambda} \sim 4-5$ MeV \cite{exper}.

Wether this additional softening is realistic or not will depend on the details 
of the $YY$ interaction that is, unfortunatelly, not well constrained at present.
New data in the $S=-2$ sector, either from double--$\Lambda$ hypernuclei or from
$\Xi^-$--atoms, are very much awaited for. 

\section*{Acknowledgements}
We are much indebted to M.~Baldo, F.~Burgio, \O.~Elgar\o y, 
H.~Heiselberg, H.-J.~Schulze 
and V.~G.~J.~Stoks for many useful comments.
This work has been supported by the DGICYT (Spain) Grant PB95-1249, 
the Program SCR98-11 from the Generalitat de Catalunya and
the Research Council of Norway. One of the authors
(I.V.) wishes to acknowledge support from a doctoral fellowship of the
Ministerio de Educaci\'on y Cultura (Spain).



\newpage

\newpage

\newpage

\newpage

\newpage

\end{document}